\documentstyle[12pt]{article} 
\newcommand{\be}{\begin{equation}} 
\newcommand{\en}{\end{equation}}
\newcommand{\bea}{\begin{eqnarray}}
\newcommand{\ena}{\end{eqnarray}}

\newcommand{\hbo}{\hbox to 1 true cm {\hfill } }

\newcommand{\vp}{\vec{p}}
\newcommand{\vq}{\vec{q}}
\newcommand{\vk}{\vec{k}}
\newcommand{\th}{ \hbox{ tanh}\left( \frac{E}{2T} \right) }

\input psfig 
\def\dslash{\partial\kern-.5em\slash}
\def\kslash{k\kern-.5em\slash}
\def\pslash{p\kern-.5em\slash}

\begin{document} 
\vglue 1truecm
  
\vbox{ UNIT\"U-THEP-1/1996 
\hfill December 22, 1995
}
\vbox{ hep-ph/9601223 } 
  
\vfil
\centerline{\bf Restoration of chiral symmetry in quark models with } 
\centerline{\bf    effective one gluon exchange$^1$ } 
  
\bigskip
\centerline{ K.\ Langfeld, H.\ Reinhardt } 
\vspace{1 true cm} 
\centerline{ Institut f\"ur theoretische Physik, Universit\"at 
   T\"ubingen }
\centerline{D--72076 T\"ubingen, Germany.}
\bigskip
  
\vfil
\begin{abstract}
  
The restoration of chiral symmetry at finite density and/or 
temperature is investigated in a wide class of one-gluon exchange models 
in the instantaneous approximation. If the effective quark 
interaction is less divergent than $1/k^2$ for small momentum transfer 
$k$, we obtain Gaussian critical exponents for the chiral phase 
transitions at finite temperature and density, respectively. 
In the opposite case, for an interaction diverging faster than 
$1/k^2$ in the infrared region, a qualitative different behavior of the 
quark self-energy near the critical Fermi momentum $k_c$ and the critical 
temperature $T_c$, respectively, is observed. In the first scenario, we 
find $k_c \approx 2 \ln 2 \; T_c$, which compares well with recent data 
from QCD lattice simulations. 

\end{abstract}

\vfil
\hrule width 5truecm
\vskip .2truecm
\begin{quote} 
$^1$ Supported by DFG. 
\end{quote}
\eject
{\it 1. Introduction } 

The properties of hadrons in a hot and/or dense medium are of general 
interest, since they can be experimentally tested in heavy ion 
collisions. These experiments might provide informations of the widely 
unknown low energy sector of QCD. Furthermore, a detailed understanding of 
the quark ground state at finite density is necessary for a wide range 
of applications, e.g.\ to describe 
compact star matter~\cite{rho1}. In these applications, the restoration 
of the spontaneously broken chiral symmetry is of particular interest. 
For medium values of the density compared with the critical density, 
a powerful argument was developed which predicts a scaling of 
certain hadronic observables with the density~\cite{br91}. This 
so-called BR-scaling yields phenomenological reasonable results 
for densities up to the critical density. Close to the critical values, 
physics is described by another approach; 
it was argued long ago~\cite{pi84}, that the chiral transition at finite 
temperature is governed by dimensional reduction and universality. 
In particular, four-dimensional QCD should belong 
to the universality class of the three-dimensional $\sigma $-model 
with the corresponding global symmetries~\cite{pi84}. Recently, 
Koci\'c and Kogut argued, that the presuppositions which support the 
universality argument, are not satisfied. In particular, they 
observe that the critical exponents of the chiral transition 
in the three-dimensional Gross-Neveu model are of Gaussian type, 
instead of those given by the Ising model, as predicted by the universality 
argument~\cite{ko95}. The chiral transition at finite density is less 
known, which is partly due to conceptual and practical 
difficulties of the numerical simulation at finite 
densities~\cite{bar92}. 
It is tacitly assumed that the critical value and indices of the chiral 
transition at finite density resemble those of the transition at 
finite temperature. 

In this letter, we study the chiral transition at finite density 
and/or temperature in a wide class of constituent quark models 
with effective one-gluon exchange. We will find that Gaussian 
critical exponents are the generic case, and will discuss the 
circumstances which imply a correspondence of the chiral transition 
at finite temperature and finite density.

\medskip 
{\it 2. Model description } 

Starting point of the one-gluon exchange models is the QCD Dyson-Schwinger 
equation for the quark propagator $S(p)$, which in Landau gauge and in 
Euclidean space-time reads 
\be 
i \pslash + m + \int \frac{ d^4k }{(2\pi )^4 } \; \Gamma ^a_\mu 
(p,p-k) \, S(k) \, \gamma _\nu t^a \; D_{\mu \nu }^{ab} (p-k) 
\; = \; S^{-1}(p) \; , 
\label{eq:1} 
\en 
where the Euclidean quark propagator is of the form 
\be 
S(p) \; = \; \frac{i}{ Z(p^2) \pslash \, + \, i \Sigma (p^2) } \; . 
\label{eq:1a} 
\en 
Equation (\ref{eq:1}) must be supplemented with Ans\"atze for the full 
gluon propagator 
$D_{\mu \nu }^{ab} (p-k) $ and the quark-gluon vertex function 
$\Gamma ^a_\mu (p,p-k)$. The variety of 
models~\cite{fi82,alk88,pa77,sme91,rob94}--and this is not a complete 
list--differ 
in the particular choice of $D_{\mu \nu }^{ab} $ and $\Gamma ^a_\mu $. 
These choices must match the well known behavior of the perturbative 
Green's functions at large momentum transfer, i.e. 
\be 
D_{\mu \nu }^{ab} (k) \; = \; \delta ^{ab} \, \frac{ 16 \pi ^2 }{ 11 \, 
\ln k^2 / \Lambda _{QCD}^2 } \; \frac{1}{k^2} \; \left( 
\delta _{\mu \nu } - \frac{ k_\mu k_\nu }{k^2} \right) \; , 
\hbox to 2 cm {\hfil for \hfil } k^2 \gg \Lambda _{QCD}^2 \; . 
\label{eq:2} 
\en 
Sometimes, the low-energy behavior of the gluon-propagator is chosen as 
\be 
D_{\mu \nu }^{ab} (k) \; \propto \; \delta ^{ab} \, \frac{1}{k^4} 
\left( \delta _{\mu \nu } - \frac{ k_\mu k_\nu }{k^2} \right) 
\hbox to 2 cm {\hfil for \hfil } k^2 \approx 0 \; . 
\label{eq:3} 
\en 
This choice produces a linear confining potential between static 
quarks. It was observed, that the ansatz (\ref{eq:3}) yields a model 
which is one-loop renormalizable and realizes confinement by 
infrared-slavery~\cite{sme91}, 
which enforces the quark propagator to vanish. 
In any case, we will assume that the interaction under consideration 
is strong enough to break chiral symmetry spontaneously. 

A different confining mechanism arises when one assumes that the 
Yang-Mills vacuum is realized by random color electro-magnetic 
fields~\cite{la95}, which induce a complex constituent quark mass. 
In this case, the quark propagator is finite, but quarks do not 
exist as asymptotic states, and furthermore the Landau pole of the 
corresponding perturbative gluon-propagator is screened due to the 
presence of a dynamical mass~\cite{la95b}. This implies that the behavior 
\be 
D_{\mu \nu }^{ab} (k) \; \propto \; \delta ^{ab} \, \frac{1}{k^2} 
\left( \delta _{\mu \nu } - \frac{ k_\mu k_\nu }{k^2} \right) 
\hbox to 2 cm {\hfil for \hfil } k^2 \approx 0 \; . 
\label{eq:4} 
\en 
is consistent with phenomenology. 

In this letter, we need not to specify the full gluon-propagator 
beyond the constraint (\ref{eq:2}), but will refer to the different 
specific scenarios (\ref{eq:3}) and (\ref{eq:4}) when needed. 

In order to stress the generality of our main result, we investigate 
to different Ans\"atze for the vertex function $\Gamma ^a_\mu $, i.e.\ 
the bare vertex 
\be 
\Gamma ^a_\mu (p,q) \; = \; t^a \gamma _\mu \; , 
\label{eq:5} 
\en 
which is the simplest choice, and a more sophisticated vertex, i.e. 
\bea 
\Gamma ^a_\mu (p,k) &=& t^a \gamma _\mu T_1(p^2,k^2) + 
(p_\mu + k_\mu ) (\kslash + \pslash ) T_2 (p^2,k^2) 
\label{eq:6} \\ 
&+& (p_\mu + k_\mu ) (\kslash - \pslash ) T_3 (p^2,k^2) 
- i (p_\mu + k_\mu ) T_4(p^2,k^2) \left[ \Sigma (k^2) - \Sigma (p^2) 
\right] \; , 
\nonumber 
\ena 
which is motivated from Taylor-Slavnov identities~\cite{sme91,cur90}. 
The last term on the right hand side of (\ref{eq:6}), which is 
proportional to the self-energy, ensures that the Ansatz does not 
explicitly break chiral symmetry. 
From dimensional arguments, one observes that the functions 
$T_{2 \ldots 4 }$ decrease like $1/p^2$ and $1/k^2$ for large values 
for $p^2$ and $k^2$, respectively. 

Using the Ansatz (\ref{eq:2}), the Dyson-Schwinger equation decays 
into two equations for $Z(p^2)$ and the quark self-energy 
$\Sigma (p^2)$. The equation for the latter can be cast into the form 
\be 
\Sigma (p^2) \; = \; m + \frac{1}{H(p^2)} \int \frac{ d^4k }{(2\pi )^4 } 
\; \frac{ \Sigma (k^2) }{ Z^2(k^2) k^2 + \Sigma ^2 (k^2) } 
\, {\cal D } (k,p) \; , 
\label{eq:7} 
\en 
where for the case of the bare vertex (\ref{eq:5}) the functions $H$ 
and ${\cal D}$ are 
\bea 
H(p^2) &=& 1 \; , 
\label{eq:8} \\ 
{\cal D } (p,k) &=& \frac{1}{2} D_{\mu \mu }^{aa} (p-k) \; , 
\label{eq:9} 
\ena 
while in the case of the more complicated vertex (\ref{eq:6}), these 
functions are 
\bea 
H(p^2) &=& 1 \, + \, \int \frac{ d^4k }{(2\pi )^4 } \; 
\frac{Z(k^2)}{ Z^2(k^2) k^2 + \Sigma ^2(k^2) } \, {\cal D} _1 (p,k) 
\label{eq:10} \\ 
{\cal D } (p,k) &=& \left[ Z(k^2) T_4(p^2,k^2) +2 T_2 (p^2,k^2) \right] 
p_\mu D^{aa}_{\mu \nu } p_\nu + \frac{1}{2} T_1(p^2,k^2) 
D_{\mu \mu } ^{aa} \; , 
\label{eq:11} 
\ena 
where 
\be 
{\cal D}_1 (p,k) \; = \; T_4(p^2,k^2) \, p_\mu D_{\mu \nu }^{aa} 
(p-k) p_\nu \; . 
\label{eq:11a} 
\en 
Note that the function $H(p^2)$ in (\ref{eq:10}) is ultra-violet finite, 
since ${\cal D}_1 (p,k)$ decreases like $1/k^4$ for large $k$. 

In order to address the renormalization of the Dyson-equation, 
we exploit the fact that the self-energy $\Sigma (k^2)$ and 
the gluon induced interaction ${\cal D }$ match the perturbative 
behavior at large momentum transfer, i.e. 
\be 
\Sigma (k^2) \; \approx \; \frac{ const. }{ \left[ 
\ln k^2/ \Lambda _{QCD}^2 \right] ^{d_m} } \; , \hbo 
{\cal D } \; \approx \; \frac{ const. }{ k^2 \, 
\ln k^2/\Lambda _{QCD}^2 } \; , 
\label{eq:12} 
\en 
where $d_m >0 $ is the anomalous dimension of the quark mass. 
The generic behavior of $Z(k^2)$ turns out to be weakly 
momentum dependent and asymptotically approaches a constant for large 
$k^2$. In fact, $Z(\mu ^2)$ is renormalized to $1$ at some large 
subtraction point $\mu $ in perturbative QCD with mass shell 
renormalization~\cite{yn83}. 
This large momentum behavior of $Z(k^2)$ and $\Sigma (k^2)$ implies 
that the momentum integration in the Dyson-equation 
(\ref{eq:7}) is ultra-violate finite. If one nevertheless cuts off 
this momentum integration by an O(4)-invariant cutoff $\Lambda $, 
the behavior of a self-consistent solution of (\ref{eq:7}) 
enforces the bare quark mass $m$ to vanish for $\Lambda 
\rightarrow \infty $ as (see e.g. \cite{co89}) 
\be 
m (\Lambda ^2) \; = \; \frac{1}{ \left[ \ln \Lambda ^2/ \mu ^2 \right] 
^{d_m} } \, m_{R}(\mu ) \; , 
\label{eq:13} 
\en 
where $\mu $ is the renormalization point and $m_R$ is the renormalized 
quark current mass. Here we are only interested in the chiral limit 
implying 
\be 
m _{R} (\mu ) =0 \; \rightarrow \; m \equiv 0 
\label{eq:14} 
\en 
from the very beginning. 

In the following, we will adopt further simplifications which we 
believe not to change our conclusions. We will assume that the function 
$Z(k^2)$ is weakly dependent on the momentum $k^2$ for the models 
which we consider here, implying that we may safely approximate 
$Z(k^2)=1$. In fact, one realizes that $Z(k^2)=1$ is a solution 
of the Dyson-equation, if one neglects vector-contributions to the 
Dyson-Schwinger equation after a Fierz rearrangement of the interaction. 
 
Finally, we assume a separation of the fermionic and gluonic 
energy scales $E_f$ and $E_g$, i.e. $E_g \gg E_f$. 
This assumption is motivated by lattice results~\cite{gro94}, which show 
that the lowest-lying excitations of the Yang-Mills vacuum, the glue balls, 
have energies larger than $\approx 1.5 \, $GeV, while the energy scale of 
the lowest quark modes is set by the constituent quark mass 
of $\approx 300 \, $MeV. 
The above presupposition has two immediate consequences; firstly, the 
contribution of the temperature dependent gluon exchange to the loop 
integration in the Dyson-Schwinger equation is of order 
$\exp \{-E_g/T \}$, which is small compared with the fermionic part 
$\exp \{-E_f/T \}$. This implies 
that we may neglect the temperature dependence of the interaction. 
In fact, this idea is supported by simulations of lattice QCD, where 
one observes that the gluon condensate is weakly temperature dependent 
throughout the chiral transition~\cite{bar92} suggesting that the chiral 
transition is solely driven by the temperature dependence of the quark 
loop. Secondly, the dominant contribution of the $k_0$ integration in the 
loop integral of the Dyson-Schwinger equation stems from the region 
$k_0 \approx E_f$\footnote{ This argument becomes obvious, if one  
performs this integration in the complex plane by Cauchy's theorem.}. 
This implies that we may neglect the $k_0$ dependence in the 
gluon exchange, and we end up with an instantaneous approximation. 

\medskip 
{\it 3. Restoration of chiral symmetry at finite density } 

In order to investigate density effects, we introduce a chemical potential 
$\mu $, which enforces a non-vanishing vacuum density 
$\rho = \langle \Omega \vert q^{\dagger } q \vert \Omega \rangle $. 
In instantaneous approximation, the quark self-energy $\Sigma $ only 
depends on the three momentum $\vec{k}$, implying that the 
$k_0$-integration in the Dyson-Schwinger equation (\ref{eq:7}) and 
in $H (p^2)$ (\ref{eq:10}) can be done, i.e. 
\bea 
\Sigma (\vp^2) &=& \frac{1}{H_f(\vp^2)} \int _{k^2 \ge k_f^2} 
\frac{ d^3k }{(2\pi )^3 } 
\; \frac{ \Sigma (\vk^2) }{ 2 \sqrt{ \vk^2 + \Sigma ^2 (\vk^2)} } 
\, {\cal D } (\vk,\vp) \; , 
\label{eq:15} \\ 
H_f(\vp^2) &=& 1 \, + \, \int _{k^2 \ge k_f^2} 
\frac{ d^3k }{(2\pi )^3 } 
\frac{1}{ 2 \sqrt{ \vk^2 + \Sigma ^2(\vk^2)} } \, {\cal D}_1 (\vp,\vk) \; , 
\label{eq:16} 
\ena 
where the subscript ``$\rho $'' indicates the finite density case. 
Furthermore, the Fermi momentum $k_f$ is related to the density by 
$k_f ^3 = \pi^2 \rho $ for $N_c=3$ colors and $N_f=1$ flavor. In the 
case of the bare vertex (\ref{eq:5}), 
the function $H_f$ is again $H_f(\vp^2)=1$ and thus independent of the Fermi 
momentum. 

We first derive an equation for the critical Fermi momentum 
$k_c$, at which chiral symmetry is restored. To this aim, we apply the 
so-called bifurcation method~\cite{atk87} and exploit the fact that 
at the critical density, a second solution to the 
Dyson-Schwinger equation (\ref{eq:15}) exists besides the trivial 
one $\Sigma \equiv 0$. At the critical density, the second solution 
differs from the trivial one only by an tiny amount implying 
that its momentum dependence $f(\vp^2)$, where $\Sigma (\vp ^2) \propto 
f(\vp^2) $ with $f(0)=1$, can be calculated from the linearized 
equation, i.e. 
\be 
f (\vp ^2) \; = \; \frac{1}{ H_0(\vp^2) } \int _{k^2 \ge k_c^2} 
\frac{ d^3k }{(2\pi )^3 } \; 
\frac{ 1 }{ 2 \sqrt{ \vk^2 } } 
\, {\cal D } (\vk,\vp) \; f(\vk^2) , 
\label{eq:17} 
\en 
where 
\be
H_0(\vp^2) \; = \; 1 \, + \, \int _{k^2 \ge k_c^2} 
\frac{ d^3k }{(2\pi )^3 } 
\frac{1}{ 2 \sqrt{ \vk^2 } } \, {\cal D}_1 (\vp,\vk) \; . 
\label{eq:18} 
\en 
This equation is essentially the Bethe-Salpeter equation describing 
fluctuations \break around the trivial vacuum at the critical point. 
Since the linearized equation is homogeneous in $\Sigma $, it does not 
fix the magnitude of the self-energy. The critical density is obtained 
from (\ref{eq:17}) by demanding that $f(\vp ^2)$ is an eigenvector of 
the integral-kernel on the right hand side of (\ref{eq:17}) with 
eigenvalue $1$. 

In order to fix the size of the self-energy for a Fermi-momentum close 
to $k_c$, we study the next to leading order of the small amplitude 
expansion of the Dyson-Schwinger equation (\ref{eq:15}), i.e. 
\bea 
\Sigma (\vp ^2) &=& \frac{1}{ H_0(\vp^2) } \int _{k^2 \ge k_f^2} 
\frac{ d^3k }{(2\pi )^3 } 
\frac{\Sigma (\vk ^2)}{ 2 \sqrt{ \vk^2 } } \, {\cal D} (\vp ,\vk ) 
\label{eq:19} \\ 
&+& \frac{1}{ 2 H^2_0(\vp^2) } \int _{k^2 \ge k_f^2} 
\frac{ d^3k }{(2\pi )^3 } 
\frac{\Sigma (\vk ^2)}{ 2 \sqrt{ \vk^2 } } \, {\cal D} (\vp ,\vk ) 
\int _{q^2 \ge k_f^2} \frac{ d^3q }{(2\pi )^3 } 
\frac{\Sigma ^2(\vq ^2)}{ 2  (\vq^2)^{3/2} } \, {\cal D}_1 (\vp ,\vk ) 
\nonumber \\ 
&-& \frac{1}{ 2 H_0(\vp^2) } \int _{k^2 \ge k_f^2} 
\frac{ d^3k }{(2\pi )^3 } 
\frac{\Sigma ^3 (\vk ^2)}{ 2 (\vk^2)^{3/2} } \, {\cal D} (\vp ,\vk ) \; . 
\nonumber 
\ena 
We will solve this integral equation approximately. On first realizes that 
for large external momentum $\vp ^2 $ the last two terms of (\ref{eq:19}) 
are irrelevant, since the functions ${\cal D} _{(1)} (\vp, \vk )$ 
are strongly peaked for 
$\vk ^2 \approx \vp ^2$ implying that the these last two terms are 
suppressed by a factor $\Sigma ^2 /\vp ^2$ compared with the first 
term in (\ref{eq:19}). We therefore conclude that for $p^2 \gg k_c^2$, 
the momentum dependence of the solution $\Sigma (\vp ^2)$ of (\ref{eq:19}) 
is given by $f(\vp ^2)$ from (\ref{eq:17}). This motivates the 
approximation 
\be 
\Sigma (\vp ^2) \; \approx \; \Sigma _{(1)}(k_f) \; f(\vp^2) \; , 
\label{eq:20} 
\en 
where $\Sigma _{(1)}$ must be calculated from (\ref{eq:19}). 
The procedure implies that the ansatz (\ref{eq:20}) is a good 
approximation to the full solution of (\ref{eq:19}) at least at 
small and at large momentum $\vp ^2$. Using the implicit expression 
(\ref{eq:17}) for $k_c$, one obtains to leading order in $k_c^2-k_f^2$ 
from (\ref{eq:19}) at $p^2=0$ 
\bea 
\Sigma _{(1)}^2 && \int _{k^2 \ge k_c^2} \frac{ d^3k }{(2\pi )^3 } 
\int _{q^2 \ge k_c^2} \frac{ d^3q }{(2\pi )^3 }  \; 
\frac{ {\cal D }(0,\vk ) f^3(\vk ^2) }{ 2 (\vk ^2)^{3/2} } 
\label{eq:21} \\ 
&& \left[ \delta (\vk - \vq ) - \epsilon \frac{ f^2(\vk ^2) }{ f^2(\vk ^2) } 
\left( \frac{ \vk ^2 }{ \vq ^2 } \right)^{3/2} 
\frac{ {\cal D}_1 (0,\vk ) }{ \sqrt{ \vk ^2 } } \right]  
\; = \; \int ^{k^2 \le k_c^2} _{k^2 \ge k_c^2} \frac{ d^3k }{(2\pi )^3 } 
\; \frac{ f(\vk ^2) }{ \sqrt{ \vk ^2 } } \, {\cal D}(0,\vk) \; , 
\nonumber 
\ena 
where $\epsilon =0 $ for case of the bare vertex (\ref{eq:5}) and 
$\epsilon =1$ for the case of the improved vertex (\ref{eq:6}). 
If one assumes that the function $T_4(p^2,k^2)$ is not singular 
for $p^2 \rightarrow 0 $ (which is the generic behavior~\cite{cur90}), 
the function ${\cal D} _1 (p^2=0,k^2)$ vanishes implying that our result 
(\ref{eq:21}) is widely independent of the ansatz for the vertex function. 
The only dependence on the particular choice enters via the function 
$T_1(p^2,k^2)$ through the projected gluon propagator ${\cal D}(p,k)$.

The integral at the right hand side of (\ref{eq:21}) is 
proportional to 
\be 
\int _{k_f^2} ^{k_c^2} du \; f(u) \, \frac{1}{2} T_1(0,u) D_{\mu \mu }^{aa} 
(u) \; . 
\label{eq:22} 
\en 
We therefore obtain finally 
\be 
\Sigma _{(1)} \; = \; \left[ \frac{ k_c^2 f(k_c^2) T_1(0,k_c^2) 
D_{\mu \mu }^{aa} 
(k_c^2) }{ 4 \pi ^2 \int _{k^2 \ge k_c^2} \frac{ d^3k }{(2\pi )^3 } 
\frac{ T_1(0,\vk ^2) D_{\mu \mu }^{aa} (\vk ^2) 
f^3(\vk ^2) }{ (\vk ^2)^{3/2} } } 
\right] ^{1/2} \; \sqrt{ 1 - \frac{k_f^2}{k_c^2} } 
\label{eq:23} 
\en 
which corresponds to a Gaussian critical exponent $1/2$.

\medskip 
{\it 4. Restoration at finite temperature } 

The quark interaction of the one-gluon exchange model is 
motivated from the gluonic interaction of QCD and therefore 
temperature dependent. However, numerical simulations of lattice QCD 
indicate that the gluonic properties only vary weakly, if the quark 
sector passes the chiral transition~\cite{kan95}. We therefore neglect here 
any explicit temperature dependence of the interaction in (\ref{eq:1}), 
but study the temperatures effects in the quark sector, since those 
apparently play the dominant role at the chiral transition. 

The standard method to calculate the impact of temperature on 
Green's function is the imaginary-time formalism~\cite{ka89}, i.e.  
one confines the configuration space of the fermionic 
fields to the configurations which are anti-periodic in the Euclidean time 
direction with a periodic length $1/T$ with $T$ being the temperature. 
At finite temperature, the integration over the zeroth component of the 
Euclidean momentum in the terms of (\ref{eq:7}) is 
replaced by a discrete sum over Matsubara frequencies. In the case 
of the instantaneous approximation, this sum can be explicitly 
evaluated and yields 
\bea 
\Sigma (\vp^2) &=& \frac{1}{H_T(\vp^2)} \int 
\frac{ d^3k }{(2\pi )^3 } \; \th 
\; \frac{ \Sigma (\vk^2) }{ 2 E(\vk ^2) } 
\, {\cal D } (\vk,\vp) \; , 
\label{eq:24} \\ 
H_T(\vp^2) &=& 1 \, + \, \epsilon \int \frac{ d^3k }{(2\pi )^3 } \; \th 
\frac{1}{ 2 E(\vk ^2) } \, {\cal D}_1 (\vp,\vk) \; , 
\label{eq:25} 
\ena 
where the subscript ``$T$'' indicates the finite temperature case, 
where $E(\vk ^2) = \sqrt{ \vk ^2 + \Sigma (\vk ^2) } $ and again 
$\epsilon =0 $ for the case of the bare vertex (\ref{eq:5}) and 
$\epsilon =1 $ otherwise. The close correspondence of the density and 
temperature dependence of the model stems from the similarity of the 
equations (\ref{eq:24}-\ref{eq:25}) and (\ref{eq:15}-\ref{eq:16}) 
respectively; whereas the $tanh $ in (\ref{eq:24}-\ref{eq:25}) 
rapidly approaches $1$ for $E \gg T$, small momenta at $E \ll T$ are 
suppressed. The same behavior is observed in the kernel of the 
integral equations (\ref{eq:15}-\ref{eq:16}) in the case of finite 
density. The only difference 
is that in this case the screening of low momenta is done 
by a step function due to Pauli blocking, whereas low momenta in the 
temperature case are linearly suppressed. 

In order to investigate the variation of the self-energy with the 
temperature near the critical temperature $T_c$, we adopt the same 
method as in the case of finite density in the previous section. We 
only sketch the calculation. 
The critical temperature and the momentum dependence of the self-energy 
is obtained from the linearized equation, i.e. 
\be 
f_T (\vp ^2) \; = \; \frac{1}{ H_0(\vp^2) } \int \frac{ d^3k }{(2\pi )^3 } 
\; \hbox{tanh} \left( \frac{ \sqrt{\vk ^2 } }{2T_c} \right) 
\frac{ 1 }{ 2 \sqrt{ \vk^2 } } \, {\cal D } (\vk,\vp) \; f_T(\vk^2) , 
\label{eq:26} 
\en 
Analogous to the density case, we approximate the self-energy for 
temperatures close to $T_c$ by 
\be 
\Sigma (\vp ^2) \; \approx \; \Sigma _{(1T)}(T_c) \; f_T(\vp^2) \; , 
\label{eq:27} 
\en 
and obtain from the to next to leading order linearized 
Dyson-Schwinger equation 
\newpage 
\bea  
\Sigma _{(1T)} &=& \left[ 4 \int \frac{ d^3k }{(2\pi )^3 } 
\frac{ e^{- k/T_c } }{ (1+ e^{- k/T_c } )^2 } f_T(\vk ^2) 
T_1 (0,\vk^2) D_{\mu \mu }^{aa}(\vk ^2) \right] ^{1/2} 
\label{eq:28} \\ 
& \times & 
\left[ \int \frac{ d^3k }{(2\pi )^3 } 
\frac{ T_c - T_c e^{ - 2 k/T_c } - 2 k e^{ -k/T_c } }{ 
k^3 ( 1 + e^{ -k /T_c } )^2 } 
T_1 (0,\vk^2) D_{\mu \mu }^{aa}(\vk ^2) f_T^3(\vk ^2) 
\right] ^{-1/2} 
\nonumber \\ 
& \times & \sqrt{ 1- \frac{T}{T_c} } \; , 
\nonumber 
\ena 
where $k := \sqrt{\vk ^2} $. 
We find a Gaussian critical exponent for the chiral phase transition 
at finite temperature.

\medskip 
{\it 5. Comparison of the phase transition at finite temperature and 
        finite density } 

All momentum integrals which occur in the previous sections, are 
ultra-violet finite, which is merely a consequence of asymptotic freedom 
and therefore independent of the details of the models under 
investigations. However, we have tacitly assumed that the integrals 
are also infra-red finite. In the case of finite density (section 3), 
this is of course true due to Pauli blocking of momentum states below 
the Fermi-momentum $k_f$. In contrast, the case of finite temperature 
(section 4) needs further studies. 

Expanding the term in the second line of (\ref{eq:28}) with respect 
to small $k$, i.e. 
\be 
\frac{ T_c - T_c e^{ - 2 k/T_c } - 2 k e^{ -k/T_c } }{ 
k^3 ( 1 + e^{ -k /T_c } )^2 } 
\; = \; \frac{1}{12 T_c^2 } \; + \; {\cal O} (k^2) \; , 
\label{eq:29} 
\en 
one observes that both integrals in (\ref{eq:28}) are infra-red finite, 
provided the product of vertex function and gluon propagator 
$T_1(0,k^2) D^{aa}_{\mu \mu } (k^2)$ does not diverge faster than 
$1/k^2$. In the derivation of (\ref{eq:28}), we must exclude 
models, which incorporate quark confinement by the gluon propagator 
via infra-red slavery (see (\ref{eq:3})). For the 
gluon-propagator (\ref{eq:3}), the infra-red divergence is screened 
by a tiny, but non-vanishing self-energy at temperatures somewhat below 
the critical one. This indicates that the restoration 
of the chiral symmetry is described by a different law than the simple 
one with critical exponent $1/2$ in (\ref{eq:28}). 
Since the restoration of the chiral symmetry at finite density is 
completely insensitive to the low energy behavior of the interaction 
due to Pauli's principle, we expect a completely different behavior 
of the self-energy at finite density and at finite temperature, respectively, 
close to the critical point, if the model's interaction is infra-red 
sensitive. In contrast, we have shown that chiral transition at finite 
density is of the same nature as the transition at finite temperature 
for interactions which are ''well behaved'' in the infra-red region. 

Finally, we search for a relation between the critical Fermi-momentum 
and the critical temperature. To this aim, we compare equations 
(\ref{eq:17}) and (\ref{eq:26}), which provide the critical Fermi momentum 
and the critical temperature, respectively. Since in the 
latter case the $\hbox{ \rm tanh } k/2T_c$ rapidly approaches $1$ for 
$k \gg T_c$, and since the integral kernel in both equation strongly 
peaks at $k \approx p$, one finds that the function $f(\vp ^2)$ 
(\ref{eq:17}) coincides with $f_T(\vp ^2)$ (\ref{eq:26}) for 
$\vp ^2 \gg T_c^2$. On the other hand, both 
functions are normalized to $1$ at $\vp ^2 =0$. We might therefore 
approximate 
\be 
f(\vp ^2) \; \approx \; f_T(\vp ^2) 
\label{eq:30} 
\en 
in the entire momentum range. Equating (\ref{eq:17}) and (\ref{eq:26}) 
at $p^2 = 0 $ then yields 
\be 
\int _{k_c}^{\infty } dk \; k \, T_1(0,k ^2) D_{\mu \mu }^{aa}(k ^2) 
f(k ^2) \; = \; 
\int _0^{\infty } dk \; k \; \hbox{tanh}\left( \frac{k}{2T_c} \right) \, 
T_1(0,k ^2) D_{\mu \mu }^{aa}(k ^2) f(k ^2) . 
\label{eq:31} 
\en 
A relation between $k_c$ and $T_c$ can be obtained, if one assumes 
that the function 
$k \, T_1(0,k ^2) D_{\mu \mu }^{aa}(k ^2) f(k ^2)$ is monotonic decreasing 
with $k$. In this case, one finds that 
\be 
\int _{k_c}^{\infty } dk \; k \, T_1(0,k ^2) D_{\mu \mu }^{aa}(k ^2) 
f(k ^2) \; \ge \; 
\int _{k_0}^{\infty } dk \; k \, T_1(0,k ^2) D_{\mu \mu }^{aa}(k ^2) 
f(k ^2) \; , 
\label{eq:32} 
\en 
where $k_0 $ must be calculated from 
\be 
\int _0 ^{\infty } dk \; \left[ \hbox{tanh}\left( \frac{k}{2T_c} \right) \, 
\; - \; \theta (k-k_0) \right] 
\; = \; 0  
\label{eq:33} 
\en 
with $\theta (k) $ the step-function. From this equation, we obtain 
$ k_0 = 2 \ln 2 \, T_c$ and therefore the final result 
\be 
k_c \; \le \; 2 \, \ln 2 \, T_c \; \approx \; 1.386 \; T_c 
\label{eq:34}
\en 
Its instructive to confront this inequality with lattice data. Recent lattice 
simulations~\cite{kan95} yield $T_c \approx 260 \, $MeV. From 
(\ref{eq:34}), we find $k_c \approx 360 \, $MeV. This value is in good 
agreement with the standard estimate~\cite{bar86} $k_c \approx m_N /3 $, 
where $m_N$ is the nucleon mass.

\medskip 
{\it 6. Conclusions } 

We have studied a wide class of effective one-gluon exchange models. 
The basic assumption is a separation of the fermionic and gluonic 
energy scales justifying an instantaneous approximation. 
The quark interaction is 
constrained in order to reproduce the well-know behavior of 
the perturbative Green's functions at high momentum transfer. 
With some further assumption, which we believe to be justified for 
most of the one-gluon exchange models, we have been able to extract 
quite detailed relations for the critical behavior of various 
quantities, like the quark self energy $\Sigma $, at the chiral 
phase transition. 
We find that $\Sigma $ as function of the Fermi momentum $k_f$ behave 
like 
\be 
\Sigma (\vp ^2=0) \; \propto \; \sqrt{ 1 - \frac{ k_f^2 }{ k_c^2} } \; , 
\label{eq:35} 
\en 
where $k_c$ is the critical value of $k_f$, and that the restoration 
of the chiral symmetry is insensitive to the 
infra-red behavior of the interaction due to Pauli blocking. 
In the case of finite temperature, the analogous behavior, i.e. 
\be 
\Sigma (\vp ^2=0) \; \propto \; \sqrt{ 1 - \frac{ T }{ T_c } } \; , 
\label{eq:36} 
\en 
is found, only if the interaction diverges weaker than $1/k^2$ 
at small momentum transfer $k$. In the opposite case of an 
''infra-red sensitive'' interaction, a completely different behavior 
of the self-energy near the critical density and the critical temperature, 
respectively, is expected. In the former case, we obtain Gaussian 
critical exponents supporting the recent ideas of Koci\`c and 
Kogut~\cite{ko95}. Assuming further that the effective quark interaction 
is a monotonic decreasing function of the momentum transfer, 
we obtain the inequality 
\be 
k_c \; \le \; 2 \ln 2 \; T_c \; , 
\label{eq:37} 
\en 
which is in good agreement with recent QCD lattice simulations.


%
%
\end{document}